\documentclass{article}
\usepackage{spconf,amsmath,graphicx}
\usepackage{bibspacing}
\usepackage{color,tabularx,multirow,pifont}
\usepackage{threeparttable}

\title{Large-Scale Learning on Overlapped Speech Detection: New Benchmark and New General System}
\name{Zhaohui Yin$^{\star\dagger}$, Jingguang Tian$^{\star}$, Xinhui Hu$^{\star}$, Xinkang Xu$^{\star}$,Yang Xiang$^{\star}$}
\address{$^{\star}$Hithink RoyalFlush AI Research Institute, Zhejiang, China \\
$^{\dagger}$College of Computer Science and Technology, Zhejiang University, China}

\begin{document}
%\ninept
%
\maketitle
\begin{abstract} 
% LSTM-tansformer models have shown promising results for speech emotion recognition (SER) tasks. 
Overlapped Speech Detection (OSD) is an important part of speech applications involving analysis of multi-party conversations. However, most of existing OSD systems are trained and evaluated on small datasets with limited application domains, which led to the robustness of them lacks benchmark for evaluation and the accuracy of them remains inadequate in realistic acoustic environments. To solve these problem, we conduct a study of large-scale learning (LSL) in OSD tasks and propose a new general OSD system named CF-OSD with LSL based on Conformer network and LSL. In our study, a large-scale test set consisting of 151h labeled speech of different styles, languages and sound-source distances is produced and used as a new benchmark for evaluating the generality of OSD systems.  Rigorous comparative experiments are designed and used to evaluate the effectiveness of LSL in OSD tasks and define the OSD model of our general OSD system. The experiment results show that LSL can significantly improve the accuracy and robustness of OSD systems, and the CF-OSD with LSL system significantly outperforms other OSD systems on our proposed benchmark. Moreover, our system has also achieved state-of-the-art performance on existing small dataset benchmarks, reaching 81.6\% and 53.8\% in the Alimeeting testset and DIHARD II evaluation set, respectively.
\end{abstract}
\begin{keywords}
overlapped speech detection, large-scale learning, deep neural network
\end{keywords}
\section{Introduction}
\label{sec:intro}
% Three Part: 1) The background of our study; 2) The importance of our study; 3) The history of our study. 4)The novation of our study. 5) Brief introduction of each section. 

Robust speech processing in realistic acoustic environments often requires overlapped speech detection (OSD)   \cite{boakye2008overlapped, charlet2013impact, diez2018but}. Due to the importance of OSD for speech processing technologies, many OSD systems have been proposed for solving this problem. However, the accuracy and robustness of OSD systems, particularly for complex scenes, remains inadequate.
 
Most previous OSD systems are supervised OSD systems based on neural networks \cite{Midia2020}. The reason is that the rapid development of deep learning has led to the performance of supervised OSD systems being much higher than that of unsupervised OSD systems based on signal processing techniques. Geiger et al. showed that LSTM+HMM is better than HMM \cite{Geiger2013}. Sajjan et al. proposed a BLSTM-based OSD system, and the proposed approach yielded improvements over a GMM-based OSD system in the augmented Multiparty Interaction (AMI) meeting corpus \cite{Sajjan2018}. 

The accuracy and robustness of supervised OSD systems based on neural networks are closely related to the sample distribution of the datasets. However, most of existing OSD systems are trained and evaluated on small datasets with limited application domains, which led to the robustness of them lacks benchmark for evaluation and the accuracy of them remains inadequate in realistic acoustic environments. In \cite{Jung2021}, the CRNN-based OSD system is only trained and evaluated in English datasets (AMI, DIHARD I, DIHARD II and VoxConverse), and it is not suitable for application scenarios in other languages, such as Chinese.  In \cite{Zheng2021}, The Transformer-based OSD system is only trained and evaluated on the meeting-style dataset (AMI and Alimeeting), and it is not suitable for other style application scenarios, such as family conversation style. In \cite{Wang2022}, The DMSNet is only trained and evaluated in the eight-channel conference style dataset (Alimeeting), and it is not suitable for single-channel application scenarios, such as telephone customer service scenarios.

  %% Figure1
  \begin{figure}
    \centering
    \includegraphics[width=0.45\textwidth]{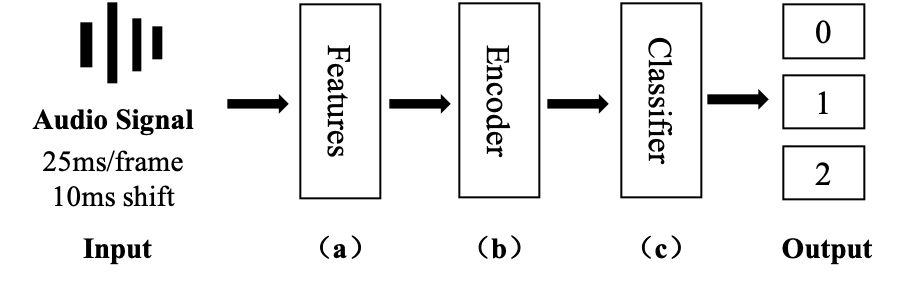}
    \caption{Flow diagram showing steps used by the OSD models with three classes: (a) Feature extraction, (b)   Encoder, (c) Classifier.}
    \label{figure1}
  \end{figure}

  \renewcommand\arraystretch{1.5}
  % % Table 1
  \begin{table*}
    \small
    \caption{The details of proposed high-quality benchmark for evaluating the generality of OSD systems and the Large-Scale Dataset for the study of Large-Scale Learning in OSD tasks.}

    \begin{tabularx}{0.98\linewidth}{c>{\centering\arraybackslash}X>{\centering\arraybackslash}X>{\centering\arraybackslash}X>{\centering\arraybackslash}Xcccccc}
      \hline
      % \makebox[0.02\textwidth][c]{Dataset} 
      % & \makebox[0.02\textwidth][c]{Subset}  
      % & \makebox[0.02\textwidth][c]{Style}
      % & \makebox[0.02\textwidth][c]{Language}
      % & \makebox[0.02\textwidth][c]{Distance}            \\
      \multirow{2}*{Dataset} &\multirow{2}*{Style}  &\multirow{2}*{Language} &\multirow{2}*{Distance} &\multirow{2}*{Clean}  
                   & \multicolumn{3}{c}{\textbf{Proposed Benchmark}}  & \multicolumn{3}{c}{\textbf{Large-Scale Dataset}}\\
                   & & & & &Subset   & \#Hours & \%Overlap  &Subset   & \#Hours & \%Overlap\\
      \hline
      MSDWILD      &Vlogs	        &Multi	  &Near &\ding{56}	&Eval   &10.58 &15.74	 &Dev  &69.09 &13.42   \\
      VoxConverse  &TV show	      &English	&Near &\ding{52}	&Test   &43.53 &3.10   &Dev  &20.30 &3.80   \\
      DIHARD 1     &Multi	        &English	&Near &\ding{52}  &Eval   &21.00 &6.06   &Dev  &19.00 &5.82   \\
      DIHARD 2     &Multi       	&English	&Near &\ding{52}  &Eval   &22.49 &6.57   &Dev  &23.81 &7.42   \\
      DIHARD 3     &Multi	        &English	&Near &\ding{52}  &Eval   &33.01 &9.35   &Dev  &34.15 &10.70   \\
      Alimeeting   &Meeting	      &Mandarin	&Far  &\ding{56}  &Test   &10.00 &18.77  &Train  &104.75 &42.27   \\
      Aishell 4    &Meeting	      &Mandarin	&Far  &\ding{56}  &Test   &12.72 &9.31   &Train  &107.50 &19.04    \\
      \hline
    \end{tabularx}
    \label{table1}
  \end{table*}
  \renewcommand\arraystretch{1.0}
As an effective tool to deal with this problem, large-scale learning has achieved remarkable results in other fields, such as computer vision, graph machine learning and so on \cite{lim2021large,reizenstein2021common}. Thus, we conduct a study of large-scale learning in OSD tasks and propose a new general OSD system named CF-OSD with LSL based on Conformer network and LSL. Our experiment is mainly divided into two parts: 1) The Production of Proposed Benchmark for evaluating the generality of OSD systems. 2) The Construction of General OSD System. The new benchmark should include labeled audio in different languages, styles and sound source distance. and these labeled audio should be collected from existing open-sourced datasets, such as DIHARD, Alimeeting, MSDWILD, and so on \cite{ryant2018first,ryant2020third,ryant2019second,chung2020spot,yu2022,fu2021aishell}. A general OSD system contains the learning method and the OSD model, thus the construction of general OSD system is to evaluate the effectiveness of LSL method and select appropriate OSD model through the comparative experiments of different OSD systems. 

The paper is organized as follows. Section 2 details the production of proposed benchmark and the construction of prposed general OSD system, Section 3 details the configuration and the result of our experiments, and Section 4 briefly summarizes our work.

\section{Methods}
\label{sec:METHODOLOGY}

  \subsection{Theoretical Basis}
  \label{ssec:theory}
  The flowchart of the OSD models is shown in Figure \ref{figure1}. As a sequence labeling task, the OSD models includes three parts: feature extraction, encoder, and classifier.

  Part I: Feature extraction. The features specifically refer to frame-level acoustic features (such as MFCC, Fbank, etc.), and they can be expressed as:
  \begin{equation}
    \textbf{X} = [\textbf{x}_1,\textbf{x}_2,....,\textbf{x}_t],\textbf{x}_t \in \textbf{R}^f
  \label{eq1}
  \end{equation}
  where $t$ is the length of input sequence and $f$ refers to the dimensionality of each element.

  Part II: Encoder. Through the encoder layer, the acoustic features $x$ can be converted into embedding $e$ which is the high-dimensional and abstract expression of features.  The embedding $e$ can be calculated by:
  \begin{equation}
    \textbf{E} = encoder(\textbf{X})
    \label{eq2}
    \end{equation}

  Part III: Classifier. The classifier is to classify the embedding obtained in the previous step. In general, the classifier includes multiple layers of linear neural network, and the output is frame-level labels $\hat{y}$. The expression of frame-level labels $\hat{y}$ is :
  \begin{equation}
    \hat{\textbf{Y}} = [\hat{y}_1,\hat{y}_2,....,\hat{y}_t],\hat{y}_t \in \textbf{R}
    \label{eq3}
    \end{equation}

  The ground truth labels $y$ provided in the dataset are used to guide the model training, and it can be expressed as eq.\ref{eq4}. 
  \begin{equation}
    \textbf{Y} = [y_1,y_2,....,y_t],y_t \in \textbf{R}
    \label{eq4}
    \end{equation}

  Different loss functions are used in the training of OSD models with different class numbers. For example, the OSD model with two classes uses the binary cross-entropy(BCE) loss function, and the OSD model with three classes uses the cross-entropy(CE) loss function.

  \subsection{The Production of Proposed Benchmark}
  \label{ssec:Dasates}
  Ideally, a general OSD system should be language independent , style independent and distance independent. Thus, The new benchmark should include labeled audio in different languages , styles and sound-source distance. 
  
  We finally collect a total of 151 hours labeled high-quality audio from seven existing open-sourced datasets, the durations of silence, single-person speech, and overlapped speech are 24h, 117h, and 10 hours, accounting for 16\%, 7\%, and 7\% respectively. More details of our proposed benchmark are shown in Table \ref{table1}.

  Table \ref{table1} shows that our new benchmark is balanced across five dimensions: style, language, sound source distance, speech cleanliness and the proportion of overlapped speech. In terms of style, proposed datasets include conferences, interviews, daily conversations, audio novels, vlogs, and many other styles. In terms of languages, considering the actual application scenarios, we selected Chinese and English as representatives. In terms of sound-source distance, our proposed datasets contains both far-field and near-field recorded speech. In terms of speech cleanliness, our proposed datasets includes clean speech, noisy speech and reverberated speech. In terms of the proportion of overlapped speech, our proposed datasets contain corpus with various levels of overlapping sound ratios.

  \subsection{The Construction of Proposed General OSD system}
  \label{ssec:selection}
  A general OSD system contains the learning method and the OSD model. Thus, the Construction of Proposed General OSD system can be divided into two parts: Evaluate the effectiveness of LSL method and Select appropriate OSD models.

  Part I : Evaluate the effectiveness of LSL method. Two comparatived OSD systems consisting of Transformer-based OSD model (TF-OSD) \cite{Zheng2021} and different learning method to evaluate the effectiveness of LSL, and we named them as the TF-OSD with LSL system and TF-OSD without LSL system. As the key of LSL, the Large-Scale Dataset should be produced first. Following the rules of benchmark production, we produced the large-scale dataset. The total duration of the large-scale dataset is 371 hours, the durations of silence, single-person speech, and overlapped speech are 38h, 281h, and 52 hours, accounting for 11\%, 75\%, and 14\% respectively. More details of large-scale dataset are also shown in Table \ref{table1}. It is worth noting that the large-scale dataset is only used to train the TF-OSD model in the TF-OSD with LSL system.

  Part II : Select appropriate OSD model. In addition to TF-OSD, we also selected three other representative models: ROSD model \cite{Sajjan2018}, TCN-based OSD model (TCN-OSD) \cite{cornell2020detecting}, and Conformer-based OSD model (CF-OSD) \cite{Wang2022}. Combined with LSL, the comparative experiments of four OSD systems are designed to select appropriate OSD model: TF-OSD with LSL, TCN-OSD with LSL,CF-OSD with LSL and ROSD with LSL. The inputs and outputs in all OSD models are the same, but the network structures are different, the details of different OSD models are shown in Table \ref{table12}.

  \renewcommand\arraystretch{1.5}
  % % Table 12
  \begin{table}[h]
  \small
  \caption{Model selection of proposed general OSD system.}
  \begin{tabularx}{0.96\linewidth}{c>{\centering\arraybackslash}Xc>{\centering\arraybackslash}X>{\centering\arraybackslash}X}
  \hline
  Layer & TF-OSD &TCN-OSD &CF-OSD &ROSD \\
  \hline
  Inputs & \multicolumn{4}{c}{Fbanks@64*400} \\
  \cline{2-5}
  Pre-Net &1*1 Conv &1*1 Conv &1*1 Conv & Liner \\
  Encoder &Transformer Encoder & TCN & Conformer Encoder & BiLSTM \\
  Post-Net & Linear &1*1 Conv &Linear &Linear \\
  \cline{2-5}
  Outputs & \multicolumn{4}{c}{Labels@3*400} \\
  \hline
  \end{tabularx}
  \label{table12}
  \end{table}
  \renewcommand\arraystretch{1.0}

  The most obvious difference among the four OSD models is the structure of the encoder. In TF-OSD model, the encoder is composed of 12 Transformer blocks, and each block contains 8-heads of attention. In TCN-OSD model, the encoder consists of 3 TCN blocks, and each TCN block contains 8 res-blocks \cite{Midia2020}. In CF-OSD model, the encoder is composed of 6 Conformer blocks, and the rest of the network settings are the same as TF-OSD. In ROSD, the encoder uses a bidirectional LSTM network. 
  
  Finally, among all OSD systems mentioned above, the one with the best performance in our proposed benchmark are defined as our new proposed general OSD system. 

\section{Experiments}
\label{sec:Experiments}
  \renewcommand\arraystretch{1.5}
  % % Table 2
  \begin{table*}[h]
    \small
    \begin{threeparttable} 
      \caption{Detailed experimental results of different OSD system}
      \begin{tabularx}{0.98\linewidth}{c>{\centering\arraybackslash}Xc>{\centering\arraybackslash}Xcccc>{\centering\arraybackslash}Xc>{\centering\arraybackslash}Xc}
        \hline
        Archi.  &Learning Method  &Aug.   &Ali-meeting\tnote{2}   &Aishell4\tnote{2}  &DH1\tnote{1}  &DH2\tnote{2}  &DH3\tnote{2} &MSD-WILD\tnote{2}  &Vox\tnote{2}  &Mean \tnote{1}   & Param.  \\
        \hline
        \multirow{2}*{TF-OSD} 
                &None	 &\ding{56}	 &73.8\%  &39.2\%   &19.5\%  &22.1\%  &26.2\%  &37.9\%  &13.8\%  &34.5\% &3.98M	\\
                &LSL	 &\ding{56}	 &78.6\%	&53.1\%   &48.0\%  &49.0\%  &56.7\%  &62.2\%  &54.9\%  &59.6\% &3.98M	\\
        TCN-OSD &LSL	 &\ding{56}	 &79.9\%	&\textbf{57.3\%}   &\textbf{50.7\%}  &\textbf{52.4\%}  &\textbf{59.3\%}  &62.9\%  &53.0\%  &60.9\% &3.87M	\\
        CF-OSD  &LSL	 &\ding{56}  &\textbf{80.8\%}	&57.0\%   &49.6\%  &52.0\%  &58.8\%  &\textbf{64.3\%}  &\textbf{55.4\%}  &\textbf{61.4\%} &4.01M	\\
        ROSD    &LSL	 &\ding{56}	 &71.7\%	&42.1\%   &34.3\%  &37.1\%  &45.4\%  &50.3\%  &35.5\%  &47.5\% &4.07M	\\
        \hline
        CF-OSD  &LSL	 &\ding{52}	 &81.6\%$\uparrow$	&58.0\% $\uparrow$  &51.0\%$\uparrow$  &53.8\%$\uparrow$  &59.1\%$\uparrow$  &65.0\%$\uparrow$  &57.5\%$\uparrow$  &62.4\%$\uparrow$ &4.01M	\\
        \hline
      \end{tabularx}
      \begin{tablenotes}
        \footnotesize 
        \item[1] Avergae F1 values of all subsets, which represents the performance of  OSD system on our new benchmark.
        \item[2] F1 values of different subset of our proposed dataset, which represents the performance of  OSD system on this specific dataset.
      \end{tablenotes}  
    \end{threeparttable} 
    \label{table2}
  \end{table*}
  \renewcommand\arraystretch{1.0}

  \subsection{Experiment Configurations}
  \label{ssec:Experiment Setup}
  We uniformly divide the speech into 4s segments, and then extract the 64-dimensional logarithmic mel spectrum (ie Filterbank) feature for each frame of the sound signal. 
  
  All models are trained with the same batch number and optimizer. There are two conditions for stopping the training of the model. One is that the loss has not decreased for six consecutive times, and the other is that the number of epochs has reached 100 times. Due to the excellent performance of Adam optimizer in other tasks \cite{dong2018speech,luo2021rethinking,he2022ustc}, we use the Adam algorithm \cite{kingma2014adam} with an initial learning rate of 0.001 to train all OSD models in this experiment. To account for the problem that fixed learning rate may skip the optimal solution, we adopt a strategy of learning rate decay during the training process, and set the decay coefficient to 0.1. 

  \subsection{Training Objective}
  \label{ssec:loss}
  Due to the proportion of overlapped speech in the dataset is much lower than the proportion of non-overlapped speech, this class imbalance problem will interfere with the training of the model. To reduce this problem, we recommend setting the labels to three classes, such as 9:1 for binary classification training, and 2:7:1 for three-class classification. 
  
  Furthermore, we give weights to categorical cross-entropy loss proportionally to further reduce the impact of the class imbalance problem, and the weighted cross-entropy loss can be expressed as eq.\ref{eq8}:
  \begin{equation}
    \mathcal{L}(x,y) = \sum_{n=1}^{N} -w_{y_n} \cdot log\frac{exp(x_{n,y_n})}{\sum_{c=1}^{C}exp(x_{n,c})} \cdot 1
    \label{eq8}
    \end{equation}
  where $x$ is the input, $y$ is the target, $w$ is the weight, $C$ is the number of classes, and $N$ spans the minibatch dimension.

  \subsection{Results}
  \label{ssec:Results}
  The baseline system for our study is the TF-OSD without LSL system \cite{Zheng2021}, which is trained on the Alimeeting dataset. The results of comparative experiments of different OSD systems are shown in Table \ref{table2} .
  
  Alimeeting testset and DH2 evaluation set are existing small dataset benchmarks, the TF-OSD without LSL system has achieved 73.8\% and 22.1\% in this two dataset, respectively. The performance of TF-OSD without LSL system on different benchmark are significantly different, indicating that the small dataset benchmark is not enough to measure the robustness and accuracy of OSD systems simultaneously. 
  
  In our new benchmark, The F1 score of TF-OSD without LSL system and TF-OSD with LSL system is 34.5\% and 59.6\%, respectively. The accuracy of the TF-OSD with LSL system if 72.7\% higher than the TF-OSD without LSL system, indicating that the LSL method can significantly improved the accuracy of OSD systems. Moreover, The F1 scores of TF-OSD with LSL system on differnet subsets of proposed benchmark are significantly improved, this result means that the LSL method can improve the robustness of OSD systems. Thus, we can get a Conclusion that the LSL method is effective in OSD tasks.

  According to the results of four different systems: TF-OSD with LSL, TCN-OSD with LSL, CF-OSD with LSL, ROSD with LSL, we can find that the CF-OSD with LSL system has the highest average F1 value of 61.4\%. The TCN-OSD with LSL and TF-OSD with LSL system were next, with 60.9\% and 59.6\% respectively. The ROSD with LSL system has the lowest average F1 value of 47.5\%. Thus, we selected the CF-OSD model as the model of our proposed general OSD system.
  
  To further improve the performance of CF-OSD with LSL system, we applied two popular augmentation methods in speech processing: additive noise and room impluse response (RIR) simulation. For additive noise, we use noisy data from the MUSAN corpus \cite{snyder2015musan}. For the room impulse response, we use the simulated RIR filter provided in \cite{ko2017study}. Noise and RIR filters are chosen randomly at each training step, and the augmentation parameters are selected by referring the recipe introduced in \cite{heo2020clova}. The result show that the average F1 value of CF-OSD with LSL system reaches 62.4\% after data augmentation, which is further improved by 1.6\%. Moreover, the CF-OSD with LSL establishes a state-of-the-art performance with an F1-score of 81.6\% and 53.8\% on Alimeeting test set and DIHARD II evaluation set, which is 4.0\% and 24.0\% higher than that of the SOTA system proposed in \cite{tian2022royalflush} and \cite{Jung2021}, respectively. This means the CF-OSD with LSL system which we proposed is the best 16k single-channel OSD system currently.

\section{Conclusions}
\label{sec:Conclusions}
In this paper, we propose a high-quality dataset as a new benchmark for evaluating the generality of OSD systems, and develop a new general 16k single-channel OSD system named CF-OSD with LSL based on Conformer network and LSL.

Our experiments show that CF-OSD with LSL system significantly outperforms other OSD systems on our proposed dataset, thus providing a powerful method for OSD tasks in realistic acoustic environments. Moreover, our system has also achieved state-of-the-art performance on existing small dataset benchmarks, reaching 81.6\% and 53.8\% in the Alimeeting test set and DIHARD II evaluation set, respectively. We hope that our contribution will provide researchers with new research avenues for OSD tasks, as well as better tools for testing models and evaluating the utility of new techniques.

More work can be done in the future. We are currently performing OSD tasks based on artificially extracted features (such as fbank, MFCC, etc.). With the development of large model in the speech field, we can use pre-trained large models to extract more accurate acoustic features and further improve the performance of the OSD model. Since there are many types of large models in recent years, such as HuBERT, wav2vec, wavLM, etc., evaluating the impact of these large models on the OSD task will be the focus of our future work.

% \vfill\pagebreak

\bibliographystyle{IEEEtran}
\bibliography{refs}

\end{document}